\newcommand{\bea}{\begin{eqnarray}} \newcommand{\eea}{\end{eqnarray}}
\newcommand{\bean}{\begin{eqnarray*}} \newcommand{\eean}{\end{eqnarray*}}

\newcommand{\s}[1]{{\scriptscriptstyle #1}}
\newcommand{\sT}{{\s T}}
\newcommand{\nn}{\nonumber \\}
\newcommand{\nnn}{\nonumber }

\documentclass[
    ,final            
  ]
  {aipproc}

\layoutstyle{6x9}

\begin{document}

\title{Gluonic Pole Matrix Elements and Universality}


\author{Leonard Gamberg}{
  address={Penn State University Berks, Department of Physics
Reading,Pennsylvania 19610-USA
}}

\author{Asmita Mukherjee}{
  address={
Indian Institute of Technology, Department of Physics, 
Powai, Mumbai 400076-India
}}

\author{Piet Mulders}{
  address={
Vrije University, Department of Physics and Astronomy, 
NL-1081 HV Amsterdam, the Netherlands}}

\begin{abstract}
We  investigate the spectral properties
of quark-quark-gluon correlators and use this to study gluonic pole 
matrix elements. Such matrix elements appear in principle both for 
distribution functions such as the Sivers function and fragmentation 
functions such as the Collins function. 
We find that for a large class of spectator models,
 the contribution of the gluonic pole matrix element for
fragmentation functions vanishes.
This result is important in the  study of universality 
for fragmentation functions.
\end{abstract}

\maketitle


\section{Introduction}
In high-energy scattering processes the structure of hadrons is 
accounted for using quark and gluon correlators; forward matrix elements
of non-local quark and gluon operators between hadronic states. 
Making an expansion in the (inverse) hard scale, the 
leading dynamical effects come from two-field
configurations at two light-like separated points,
which are easily interpreted as parton densities or parton decay
functions~\cite{Collins:1981uw,Jaffe:1991ra}. These are the parton
distribution functions depending on the momentum fraction $x$ relating 
the parton momentum $k = x\,P$ to the hadron momentum $P$
or the fragmentation functions of partons into hadrons depending
on the momentum fraction $z$, relating the parton momentum $k$ and
the hadron momentum $P = z\,k$.    At sub-leading order in the hard scale or 
when explicitly measuring  transverse momenta, other matrix elements become 
important such as the three-parton correlators containing parton fields
at three different space-time points with light-like separations
and two-parton correlators with also transverse separation (light-front
correlations). These latter (light-front) correlators are described
in terms of transverse momentum dependent (TMD) distribution and 
fragmentation functions, 
which are  sensitive to the intrinsic transverse momenta
of partons in hadrons, $k = x\,P + k_\sT$ in a frame in which the
hadron does not have transverse momentum ($P_\sT = 0$) or for
fragmentation $k = \frac{1}{z}\,P + k_\sT$. In this case one
often refers to the hadron transverse momentum $P_\perp = -z\,k_\sT$ 
(in a frame in which the parton does not have a transverse momentum
($k_\perp = 0$)). 

Here, we  investigate multi-parton
correlators with one additional gluon in which the zero-momentum
limit will be studied~\cite{Efremov:1981sh,Qiu:1991pp}. 
These are so-called gluonic pole matrix
elements or Qiu-Sterman matrix elements, that have opposite 
time-reversal (T) behavior as compared to
the matrix elements without the gluon. Such matrix elements involving 
time-reversal odd (T-odd) operator combinations are of interest because
they are essential for understanding single spin asymmetries 
in high energy scattering processes 
e.g. semi-inclusive deep inelastic scattering (SIDIS) 
and Drell-Yan scattering.  In the collinear case T-symmetry can be used 
as a constraint on the parton correlators, limiting the distribution functions (DFs) to 
T-even ones. This  constraint does not apply for the fragmentation correlator 
because the final state hadron is part of a jet 
and as such is not a plane wave, 
allowing both T-even and T-odd fragmentation functions (FFs). Including 
transverse momentum dependence, both the distribution and
fragmentation correlators ($\Phi$ and $\Delta$) are
no longer constrained by T-symmetry. The reason is that the
appropriate color gauge invariant operators in the correlator,
are not T-invariant.  The T-odd operator structure
can be traced back to the color gauge link that necessarily appears in
correlators to render them color gauge-invariant. But the operator
structure of the correlator is also a consequence of the necessary
resummation of all contributions that arise from collinear gluon
polarizations, i.e.\ those along the hadron momentum. 
How this  resummation takes effect is a matter of 
calculation~\cite{Bomhof:2006ra}.  The result is a 
process dependence in the path in the gauge link.  After azimuthal weighting of cross sections one 
simply finds that the T-odd features originating from the gauge link
lead to specific factors with which the T-odd functions appear in 
observables. For DFs this provides a mechanism leading to T-odd
functions, such as the Sivers function~\cite{Sivers:1989cc}.
Comparing T-odd effects in DFs in
semi-inclusive deep inelastic scattering (SIDIS) and the Drell-Yan 
process one finds a relative minus sign~~\cite{Collins:2002kn,Brodsky:2002rv}. 
Similarly, comparing T-odd effects in FFs in SIDIS and electron-positron
annihilation one also finds a relative minus sign, at least for the
T-odd effect originating from the operator structure (gauge link)~\cite{Boer:2003cm}.
But, for FFs there are now
in principle two mechanisms leading to T-odd functions~\cite{Boer:2003cm}. 
However, the two mechanisms leading to 
T-odd functions can be distinguished.  The effect coming from the 
hadron-jet final state not being a plane
wave will not lead to process dependent factors.

In order to understand the basic features of these matrix elements we 
perform a spectral analysis by modeling the distribution and
fragmentation functions under {\em reasonable} 
assumptions~\cite{Gamberg:2008yt}.  
In particular we consider the differences between distribution
and fragmentation functions using a spectral analysis  
while restricting the momentum dependence and
asymptotic behavior of the vertices.   
In this context, the relevant gluonic pole matrix elements 
that we  study~\cite{Gamberg:2008yt} are 
$\Phi_G(k,k-k_1)$ and $\Delta_G(k,k-k_1)$ shown in 
Figs.~\ref{dis} and \ref{frag}. Of these matrix
elements only the dependence on the collinear components
$x$ and $x_1$ in the expansion of the momenta are needed
(note, the gluon momentum is parameterized as
$k_1 =[k_1^-,x_1, k_{1\sT}]$ in these figures).
We find that while both $\Phi_G(x,x-x_1)$
and $\Delta_G(x,x-x_1)$ are nonzero,
taking the limit $x_1 \rightarrow x$, $\Phi_G(x,x)$ 
remains non-zero, while $\Delta_G(x,x)$ vanishes.  
The vanishing of the T-odd gluonic pole matrix elements is important
in the study of universality of  TMDPDFs and 
TMDFFs.

\section{Quark-Gluon Correlators and Gluonic Poles}
\begin{figure}
\centerline{\includegraphics[width=0.3\columnwidth]{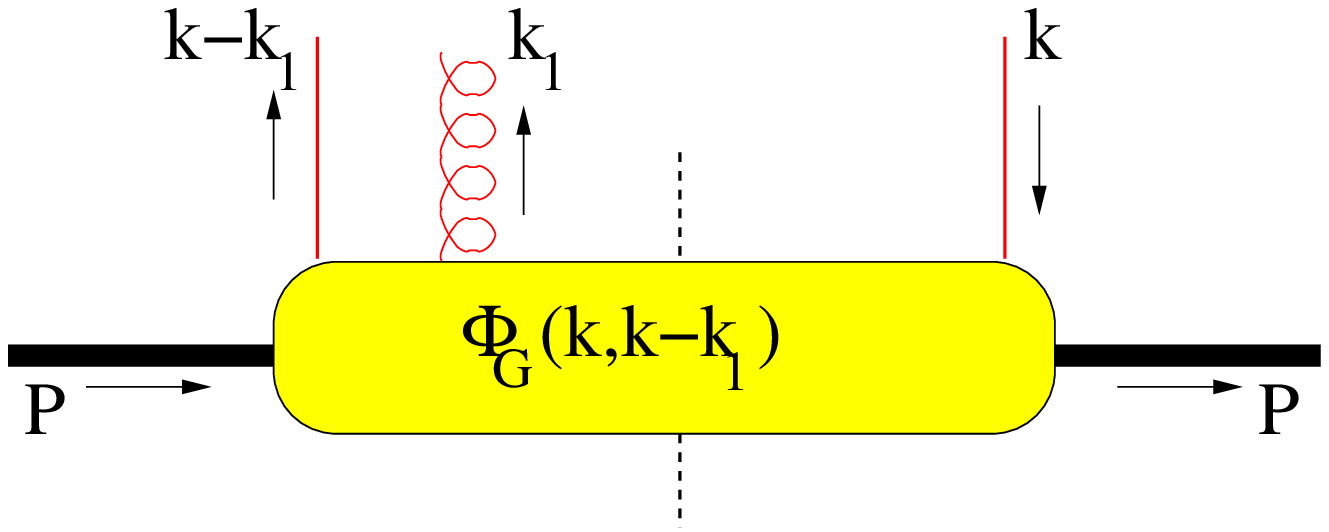}(a)
\includegraphics[width=0.3\columnwidth]{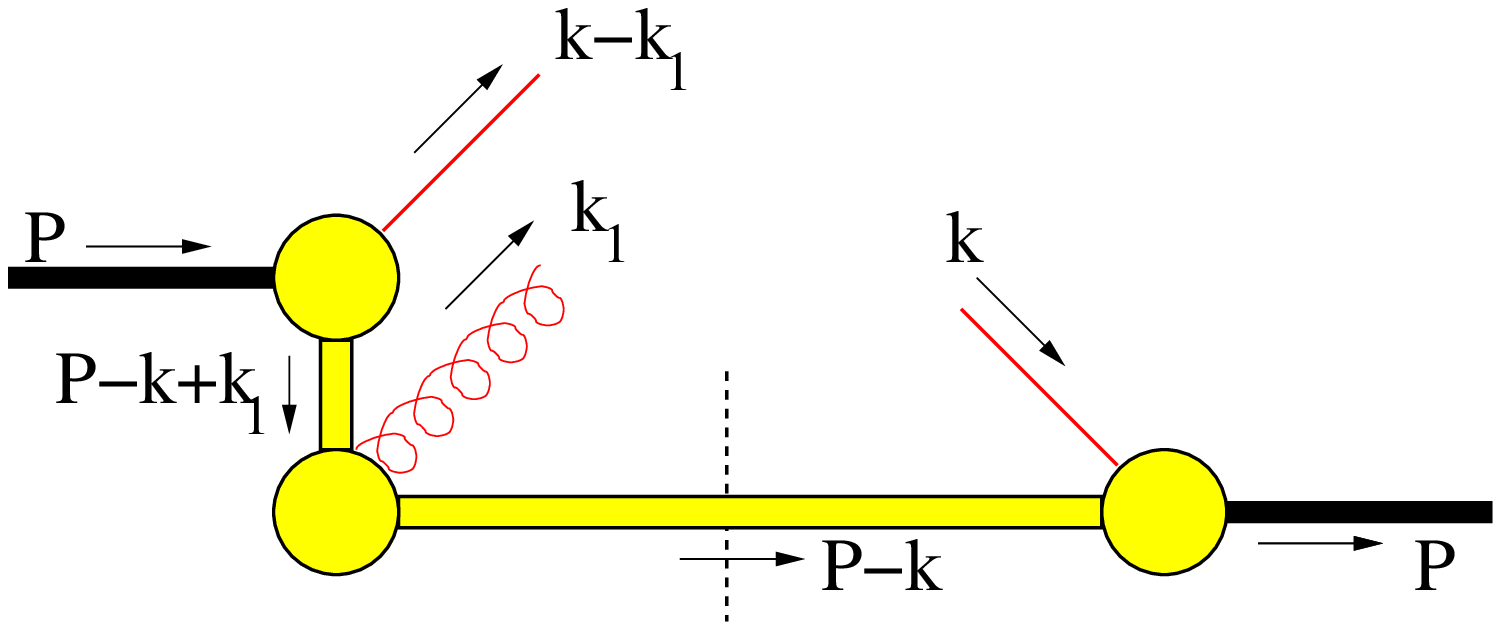}(b)
\includegraphics[width=0.3\columnwidth]{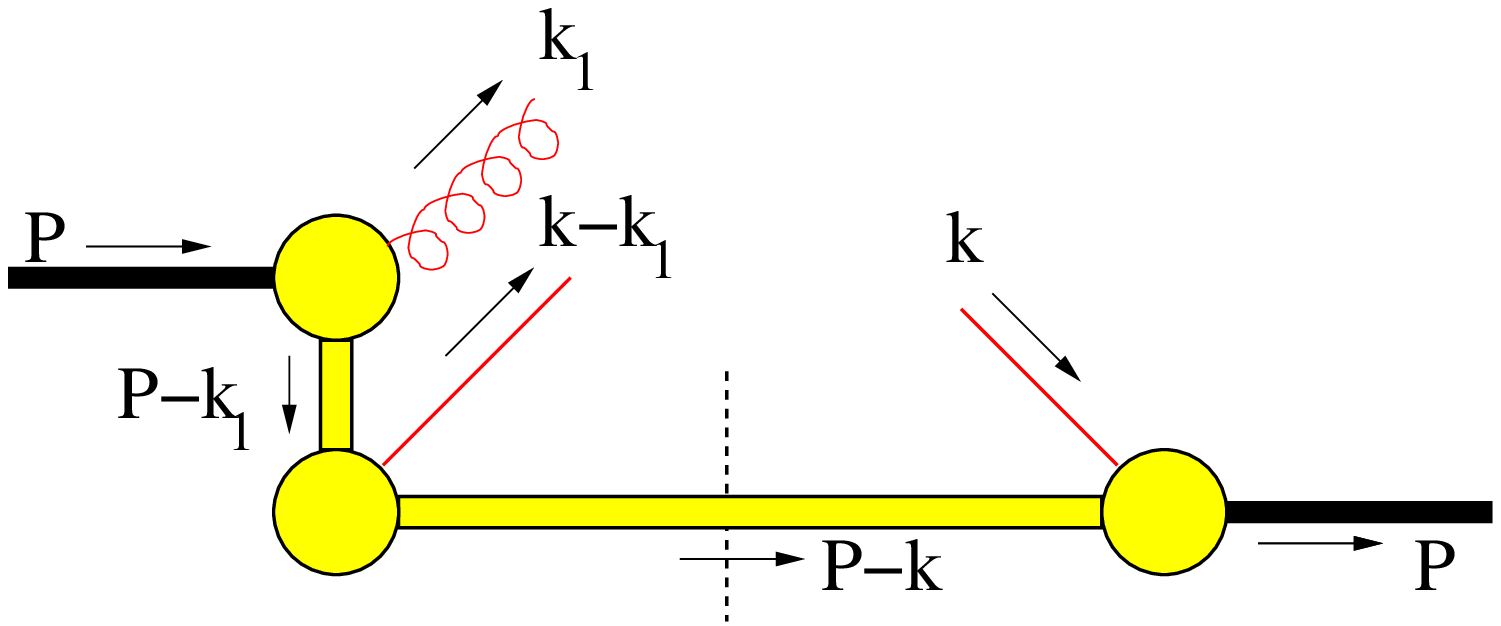}(c)}
\caption{\label{dis}
{ The graphical representation of the quark-quark-gluon
correlator $\Phi_G$ for  the case of distributions including
a gluon with momentum $k_1$ (a), and the
possible intermediate states (b) and (c) 
in a spectator 
model description. Conjugate contribution to (b) and (c)  not shown.}}
\end{figure}
The quark-quark correlator
depending on the collinear and transverse components
of the quark momentum, $k=x\,P+\sigma\,n+k_\sT$, where the
Sudakov vector $n$ is an arbitrary light-like four-vector $n^2=0$ 
that has non-zero overlap $P\cdot n$ with the hadron's momentum $P$
 and 
$k^-\sim\sigma$ (which is suppressed w/r to the hard scale) 
 is given by,
\bea\label{TMDDF}
\Phi_{ij}^{[\mathcal U]}(x{,}k_\sT)
={\int}\frac{d(\xi{\cdot}P)\,d^2\xi_\sT}{(2\pi)^3}\ e^{ik\cdot\xi}
\langle P|\,\overline\psi_j(0)\,\mathcal U_{[0;\xi]}\,
\psi_i(\xi)\,|P\rangle\big\rfloor_{\rm{LF}}\ ,
\eea
where LF ($\xi\cdot n=0$) designates the light-front. 
The  \emph{gauge link} is the path-ordered exponential, 
$\mathcal U_{[\eta;\xi]}
=\mathcal P{\exp}\big[{-}ig{\int_C}\,ds{\cdot}A^a(s)\,t^a\,\big]$
 along the integration path $C$ with
endpoints at $\eta$ and $\xi$.
Its presence in the hadronic matrix element is required by gauge-invariance.
In the correlator the integration path $C$ in
the gauge link designates  process-dependence. 
This is due to the observation
that the operator
structure of the correlator is also a consequence of the necessary
resummation of all contributions that arise from collinear gluon
polarizations, i.e.\ those along the hadron momentum.   Collinear 
quark distribution functions are obtained from the TMD correlator
after integration over $p_\sT$,
\bea\label{colcorrelator}
\Phi(x)&=& \int d^2k_\sT\ \Phi^{[\mathcal U]}(x{,}k_\sT)
={\int}\frac{d(\xi{\cdot}P)}{2\pi}\ e^{i\,x\,\xi\cdot P}\,
\langle P|\,\overline\psi(0)\,\mathcal U_{[0;\xi]}^n\,
\psi(\xi)\,|P\rangle\big\rfloor_{\rm{LC}}\ .
\eea
The non-locality is restricted to the light-cone 
(LC: $\xi\cdot n = \xi_\sT=0$) and the gauge link is
unique, being the straight-line path along $n$. 
In azimuthal asymmetries one needs the transverse moments contained in the
correlator
\bea
\label{TransverseMoment}
\Phi_{\partial}^{\alpha\,[\mathcal U]}(x) 
= \int d^2k_\sT\ k_\sT^\alpha\,\Phi^{[\mathcal U]}(x{,}k_\sT)\, .
\eea
The TMD correlator, expanded in distribution functions depending on
$x$ and $k_\sT^2$ contains T-even and T-odd functions, since the
correlator is not T-invariant. This property is attributed to the gauge
link, that depending on the process, accounts for specific initial 
and/or final state interactions depending on the color flow in the process.
For the collinear case, the link structure becomes unique in the case
of integration over $k_\sT$ (Eq.~\ref{colcorrelator}).

For the collinear weighted case, the transverse moments in 
Eq.~\eqref{TransverseMoment} one retains a nontrivial link-dependence
that prohibits the use of T-invariance as a constraint. It is possible
however, to decompose the weighted quark correlators as
\begin{equation}\label{decomposition}
\Phi_\partial^{\alpha\,[\mathcal U]}(x)
=\tilde \Phi_\partial^{\alpha}(x)
+C_G^{[\mathcal U]}\,\pi\Phi_G^{\alpha}(x,x)\, ,
\end{equation}
with calculable process-dependent gluonic pole factors $C_G^{[\mathcal U]}$
and process (link) independent correlators $\tilde\Phi_\partial$ and
$\Phi_G$. The correlator $\tilde\Phi_\partial$ contains the T-even operator 
combination, while $\Phi_G$ contains the T-odd operator combination.
The latter is precisely the soft limit $x_1\rightarrow 0$ of
a quark-gluon correlator $\Phi_G(x,x_1)$ of the type
\bea
\label{GP}
\Phi_G^\alpha(x,x{-}x_1)
&=&n_\mu
\int\frac{d(\xi{\cdot}P)}{2\pi}\frac{d(\eta{\cdot}P)}{2\pi}\ 
e^{ix_1(\eta\cdot P)}e^{i(x-x_1)(\xi\cdot P)}\,
\nn &&\times
\langle P|\,\overline\psi(0)\,U_{[0;\eta]}^n\,gG^{\mu\alpha}(\eta)\,
U_{[\eta;\xi]}^n\,\psi(\xi)\,|P\rangle\,\big\rfloor_{\rm{LC}}\ .
\eea
The universal T-odd distribution functions in the parameterization 
of $\Phi_G(x,x)$ appear in T-odd observables such as single spin 
asymmetries with the specific gluonic pole factors from 
Eq.~\ref{decomposition}.

The situation for fragmentation functions is different. The TMD fragmentation
correlator depending on the collinear and transverse components of the
quark momentum, 
$k = \frac{1}{z}\,P + k_\sT + \sigma\,n$,
is given by~\cite{Boer:2003cm,Bomhof:2006ra}
\bea
\Delta^{[\mathcal U]}_{ij}(z,k_\sT)&=&
\sum_X\int\frac{d(\xi\cdot P_h)\,d^2\xi_\sT}{(2\pi)^3}\ e^{i\,k\cdot\xi}
\langle 0 |\mathcal U_{[0,\xi]}\psi_i(\xi)|P,X\rangle
\langle P,X|\bar{\psi}_j(0)|0\rangle |_{LF}\,  .
\label{TMDFF}
\eea
The collinear, $k_\sT$-integrated correlator
$\Delta(z)=\int d^2k_\sT
 \Delta^{[\mathcal U]}(z,k_\sT)$
only contains a T-even operator combination. Nevertheless one could
in principle have T-even and T-odd fragmentation functions depending
on $z$ since the hadronic state $\vert P,X\rangle$ is an out-state,
which is not T-invariant.  In the transverse moments obtained after $k_\sT$-weighting,
\bea
\Delta^{\alpha\,[\mathcal U]}_{\partial}(z)
&=&\int d^2k_\sT\ k_\sT^\alpha \Delta^{[\mathcal U]}(z,k_\sT)
=\tilde{\Delta}_\partial^\alpha\left(\frac{1}{z}\right)
+C_G^{[\mathcal U]}
\,\pi\Delta_G^\alpha\left(\frac{1}{z},\frac{1}{z}\right),
\label{decompfrag}
\eea
the two link independent correlators $\tilde\Delta_\partial$
and $\Delta_G$ contain again a T-even and T-odd operator combination,
respectively. The gluonic pole correlator is again the soft limit,
$z_1^{-1} = x_1 \rightarrow 0$, of the quark-gluon correlator 
\bea
\Delta_{G\,ij}^\alpha\left(x,x-x_1\right)
&=&\sum_X \int\frac{d(\xi{\cdot}P)}{2\pi}\frac{d(\eta{\cdot}P)}{2\pi}\,
e^{i\,x_1(\eta\cdot P)}e^{i\,(x-x_1)(\xi\cdot P)}\,
\nn &&\times
\langle 0 | \mathcal U^n_{[0,\eta]}\, gG^{n\alpha}(\eta)
\,\mathcal U^n_{[\eta,\xi]}\psi_i(\xi)|P,X\rangle
\langle P,X|\overline{\psi}_j(0)|0\rangle\Bigg|_{LC} .
\label{GLa}
 \eea
As stated above, 
because of the appearance of hadronic states $\vert P,X\rangle$,
each of correlators in Eq.~\ref{decompfrag} contains in principle 
T-even and T-odd functions.
Rather than having a doubling of T-odd functions, we will
show in a spectator model approach that $\Delta_G(x,x)$ = 0,
which implies that T-odd fragmentation functions in the
transverse moments only come from $\tilde\Delta_\partial$,
which appear with a universal strength (no gluonic pole factors).
We demonstrate this starting with
the collinear quark-gluon correlators, Eqs.~\ref{GP} and~\ref{GLa}
rather than the model 
approaches~\cite{Ji:2002aa,Gamberg:2003ey,Bacchetta:2003rz,Bacchetta:2007wc,Gamberg:2007wm}
that looked at the  transverse momentum dependent 
quark correlators in Eqs.~\ref{TMDDF} and \ref{TMDFF}.

\begin{figure}
\centerline{\includegraphics[width=0.225\columnwidth]{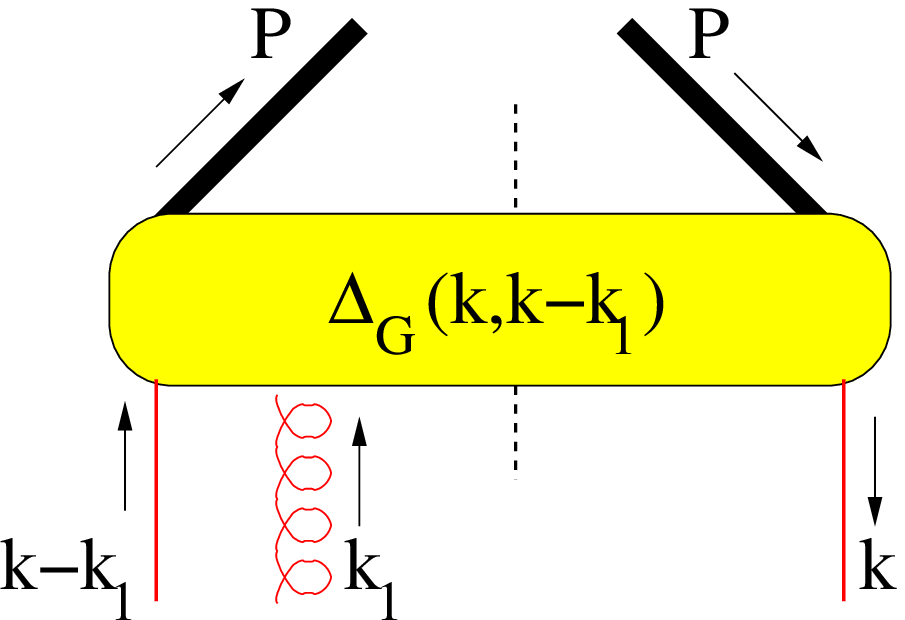}
(a)\includegraphics[width=0.3\columnwidth]{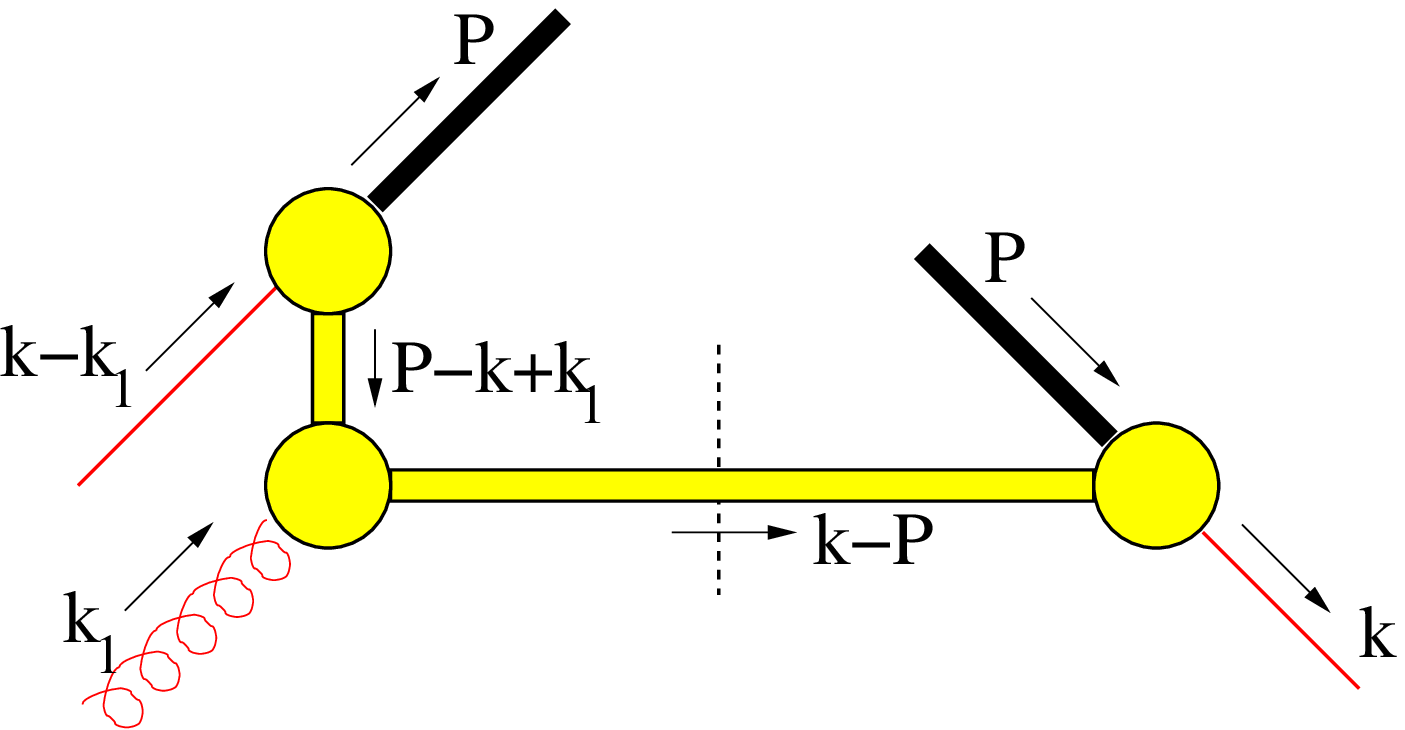}
(b)\includegraphics[width=0.3\columnwidth]{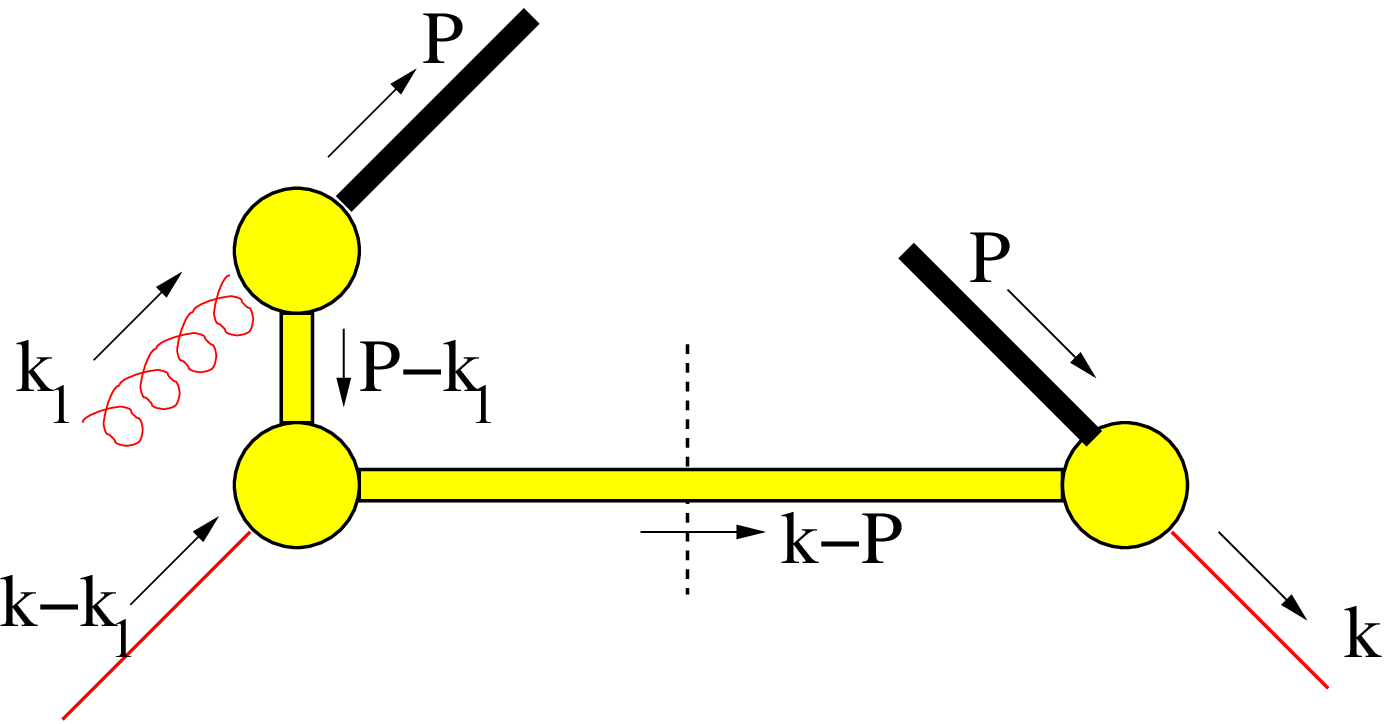}
(c)}
\caption{\label{frag}
{ The graphical representation of the quark-quark-gluon
correlator $\Delta_G$ in the case of fragmentation including
a gluon with momentum $k_1$ (a) and the
possible intermediate states (b)  in a spectator 
model description. Conjugate contribution to (b)  and (c) not shown.
}}
\end{figure}

The T-odd operator parts
are precisely the soft limits ($k_1\rightarrow 0$ or $x_1$ and
$z^-1=x_1\rightarrow 0$) 
of the gluonic pole matrix elements~\cite{Boer:2003cm} Eqs.
\ref{decomposition} and \ref{decompfrag}
(see Figs.~\ref{dis} and \ref{frag}).  
As mentioned above, they 
arise in the decomposition of   the transverse weighted quark
 correlators
which are the relevant operators in analyzing the azimuthal asymmetries. 
The process-dependent gluonic pole factors $C_G^{[\mathcal U]}$ are calculable
and  the process (link) independent correlators $\tilde\Phi_\partial$ and
 $\tilde{\Delta}_\partial$ contains the T-even operator 
combination, while $\Phi_G$ and $\Delta_G$ contain the T-odd operator combination.  The latter one is 
precisely the soft limit, 
of the quark-gluon correlator 
$\Delta_{G\,ij}^\alpha(x,x_1)$, Eq.~\ref{GLa}.

To see this in 
a spectator model approach, we consider
the  distribution or fragmentation
correlators with a spectator with mass $M_s$. The 
result for the cut, but untruncated, diagrams, such as in 
Figs.~\ref{dis} and \ref{frag} (without the gluon insertion) 
are of the form
\bea
\Phi(x,k_\sT)\hspace{-.2cm} &\sim &\hspace{-.2cm}\int d(k\cdot P)
\frac{F(k^2,k\cdot P)}{(k^2-m^2+i\epsilon)^2}
\delta\left( (k-P)^2 - M_s^2\right),
\nnn
\label{basic}
\eea
where  $F(k^2,k\cdot P)$  contains the numerators of propagators and/or
traces of them in the presence of Dirac Gamma matrices, as well as 
the vertex form factors (see for example~\cite{Jakob:1997wg}).
In the above the delta function constraint in Eq.~\ref{basic} has
been implemented. One finds that the numerator
$F(k^2,k\cdot P) = F(x,k_\sT^2)$ and hence
\bea
\Phi(x,k_\sT) \sim 
\ \frac{(1-x)^2\,F(x,k_\sT)}{\left(\mu^2(x)-k_\sT^2\right)^2},
\label{qqspec}
\eea
with $\mu^2(x) = x\,M_s^2+(1-x)\,m^2-x(1-x)\,M^2.$
Note that $k_\sT^2 = -\mathbf{k_\sT}^2 \le 0$.
The details of the numerator function depend on the details of the 
model, including the vertices, polarization sums, etc. These must
be chosen in such a way as to not produce unphysical effects,
such as a decaying proton if $M \ge m+M_s$, thus $m$ in Eq.~\ref{basic}
must represent some constituent mass in the quark propagator, rather
than the bare mass.
The useful feature of the result in Eq.~\ref{qqspec} is its ability to
produce reasonable valence and even sea quark distributions using the
freedom in the model.
The results for the fragmentation function in the spectator
model is identical upon the substitution of $x = 1/z$~\cite{Jakob:1997wg}. 

We turn to the same spectral analysis of the gluonic pole
correlator using the picture given in Figs.~\ref{dis} and~\ref{frag} for
distribution and fragmentation functions respectively.
 Again, we only need to investigate
one of the cases. Parameterizing 
the gluon momentum as 
$k_1 = [k_1^-,x_1, k_{1\sT}]$, $k_1^- = k_1\cdot P - \frac{1}{2}\,x_1\,M^2$ 
is the first component to be integrated over~\cite{Gamberg:2008yt}.
Assuming that the numerator does not grow with $k_1^-$ one can
easily perform the $k_1^-$ integrations 
assuming that the $F_i$ are independent of $k_1^-$.  
Taking the limit $x_1 \rightarrow 0$ 
of the basic result for the quark-gluon correlators 
$\Phi_G(x,x-x_1,k_\sT,k_\sT-k_{1\sT})$
we obtain  the gluonic pole correlators,
for distribution functions 
($0\le x \le 1$) (see~\cite{Gamberg:2008yt} for details)
\bea
\Phi_G(x,x) 
&=& - \int d^2k_\sT\,d^2k_{1\sT}
\ \frac{(1-x)\,F_1(x,0,k_\sT,k_{1\sT})\theta(1-x)}{
\bigl(\mu^2-k_\sT^2\bigr) \bigl(x\,B_1+(1-x)\,A_2\bigr)\,A_1}\, ,
\eea
where $A_i(\{m_i^2\},\{k_{iT}^2\},\{x_i\})$,  
and for fragmentation functions ($x = 1/z \ge 1$) 
\bea
\Delta_G(x,x) &=& 0\, .
\eea
This result depends on the assumption that the numerator
does not grow with $k_1^-$, otherwise,
 one does not get the required $x_1\,\theta(x_1)$ behavior in the
calculation~\cite{Gamberg:2008yt}. 
In  models, terms proportional to $k_1^- \sim k_1\cdot P$ 
may easily arise from numerators of fermionic 
propagators~\cite{Gamberg:2006ru}
which may easily be suppressed by  form factors at the
vertices. To prove a proper behavior within QCD one would need to study the
fully unintegrated correlators such as e.g.\ in 
Ref.~\cite{Collins:2007ph}
and show that they fall off sufficiently fast as a function of $k_1\cdot P$.
 
While our analysis is  not yet the full proof that 
gluonic pole  matrix elements vanish in the case of fragmentation,
it is a step towards such a  proof and 
the possible direction to
obtain such a proof by  considering the appropriate color gauge-invariant
soft matrix elements.
Such a proof is important as it eliminates a whole class of matrix
elements parameterized in terms of T-odd fragmentation functions besides
the T-odd fragmentation functions 
in the parameterization of the two-parton correlators.


\begin{theacknowledgments}
L.G. thanks the organizers of Diffraction 2008 for the invitation to
present this work. L.G. acknowledges 
support from  U.S. Department of Energy under 
contract DE-FG02-07ER41460.
\end{theacknowledgments}



\bibliographystyle{aipproc}   


\bibliography{gamberg_leonard_refs}

\IfFileExists{\jobname.bbl}{}
 {\typeout{}
  \typeout{******************************************}
  \typeout{** Please run "bibtex \jobname" to optain}
  \typeout{** the bibliography and then re-run LaTeX}
  \typeout{** twice to fix the references!}
  \typeout{******************************************}
  \typeout{}
 }

\end{document}